\pdfoutput=1

\documentclass[12pt]{article}

\usepackage{amsfonts}
\usepackage{amsmath}
\usepackage{amssymb}
\usepackage{mathrsfs}
\usepackage{bigints}
\usepackage{booktabs}
\usepackage[nosort]{cite}
\usepackage{color}
\usepackage{dsfont}
\usepackage{float}
\usepackage{framed}
\usepackage{graphicx}
\usepackage{indentfirst}
\usepackage{mathrsfs}
\usepackage{multirow}
\usepackage{pdflscape}
\usepackage{setspace}
\usepackage{subdepth}
\usepackage{subfig}
\usepackage{titlesec}
\usepackage[dotinlabels]{titletoc}
\usepackage{wrapfig}
\usepackage[all]{xy}
\usepackage{young}
\usepackage[vcentermath]{youngtab}
\usepackage{relsize}
\usepackage{stackengine}
\usepackage{datetime}

\usepackage{hyperref}

\numberwithin{equation}{section}

\usepackage{verbatim}

\newcommand{\be}{\begin{equation}}
\newcommand{\ee}{\end{equation}}

\newcommand{\bml}{\begin{multline}}
\newcommand{\emll}{\end{multline}}

\newcommand{\nn}{\nonumber}

\def\({\left(} \def\){\right)}
\def\[{\left[} \def\]{\right]}

\def\al{\alpha}

\def\eps{\epsilon}

\def\v{\vec}

\def\g{\gamma}

\def\lam{\lambda}

\def\d{\partial}

\def\o{\omega}

\newcommand{\la}{\langle}
\newcommand{\ra}{\rangle}

\newcommand{\bea}{\begin{eqnarray}}
\newcommand{\eea}{\end{eqnarray}}

\newcommand{\der}[2]{\frac{\partial#1}{\partial#2}}

\usepackage[left=2cm,right=2cm,top=2cm,bottom=2cm]{geometry}
\titleformat{\section}{\large\bfseries}{\thesection.}{4pt}{}
\titlespacing{\section}{0pt}{22pt}{6pt}

\titleformat{\subsection}{\normalfont\bfseries}{\thesubsection.}{4pt}{}
\titlespacing{\subsection}{0pt}{18pt}{6pt}

\titleformat{\subsubsection}{\normalfont\itshape}{\thesubsubsection.}{4pt}{}
\titlespacing{\subsubsection}{0pt}{16pt}{6pt}

\def\ie{\begin{equation}\begin{aligned}}
\def\fe{\end{aligned}\end{equation}}

\def\tilde{\widetilde}

\def\d{\partial}

\def\1{{\mathds 1}}

\def\mL{\mathcal{L}}

\def\o{\omega}
\def\v{\vec }

\DeclareFontShape{OT1}{cmr}{mx}{n}%
    {<->cmr10}{}
\newcommand{\mytitlefont}{\fontseries{mx}\selectfont}
\DeclareMathAlphabet{\titlemath}{OT1}{cmr}{mx}{n}

\newcommand{\bi}{\begin{itemize}}
\newcommand{\ei}{\end{itemize}}

\newcommand{\sss}{\subsubsection}

\usepackage{bm}

\usepackage{tikz}
\tikzset{every picture/.style={line width=0.75pt}} 

\begin{document}

\begin{titlepage}

\begin{center}

~\\[1cm]

{\fontsize{20pt}{0pt} \mytitlefont Strong wave turbulence in strongly local large $N$ theories  }\\[10pt]

~\\[0.2cm]

{\fontsize{14pt}{0pt}Vladimir Rosenhaus and Daniel Schubring}

~\\[0.1cm]

\it{ Initiative for the Theoretical Sciences}\\ \it{ The Graduate Center, CUNY}\\ \it{
 365 Fifth Ave, New York, NY 10016, USA}\\[.5cm]

~\\[0.6cm]

\end{center}

\noindent

We study wave turbulence in systems with two special properties: a large number of fields (large $N$) and a nonlinear interaction that is strongly local in momentum space. The first property allows us to find the kinetic equation at all interaction strengths -- both weak and strong, at leading order in $1/N$. The second allows us to turn the kinetic equation -- an integral equation -- into a differential equation. We find stationary solutions for the occupation number as a function of wave number, valid at all  scales. As expected, on the weak coupling end the solutions  asymptote to Kolmogorov-Zakharov scaling. On the strong coupling end, they asymptote to either the widely conjectured generalized Phillips spectrum (also known as critical balance), or a  Kolmogorov-like scaling exponent. 
\vfill 

\today
\end{titlepage}

\tableofcontents
~\\[-20pt]

\vspace{.6cm}

\section{Introduction}

Wave turbulence is a ubiquitous phenomenon in which excitations are waves and energy is transferred among scales with a constant flux \cite{Zakharov, hasselmann_1962, Falkovich, Nazarenko, FalconMordant}.  The most well known example of wave turbulence is for gravity waves on the surface of a fluid, such as  the ocean. For long wavelength waves,  the interaction and dispersion relation are scale-invariant and the nonlinearity is weak, and one can derive the scale-invariant turbulent spectrum (Kolmogorov-Zakharov scaling). For short wavelength waves, the nonlinearity becomes strong and  weak wave turbulence theory is no longer applicable. There is reason to believe that at very short scales the spectrum is again scale-invariant, but with a different exponent, referred to as the generalized Phillips spectrum  \cite{ Newell}. However, there is   limited analytic understanding of strong wave turbulence.

%Wave turbulence is a ubiquitous phenomenon in which excitations are waves and energy is transferred among scales with a constant flux. 
%In the regime in which the waves are small, so that the nonlinearity is weak, and the interaction and dispersion relation are scale-invariant, one can derive the scale-invariant turbulent spectrum (Kolmogorov-Zakharov scaling). This is weak wave turbulence \cite{Zakharov, hasselmann_1962, Falkovich, Nazarenko, FalconMordant}. A major problem is to understand the behavior of wave turbulence at strong nonlinearity. The most well known example of wave turbulence is for gravity waves on the surface of a fluid, such as  the ocean. The nonlinearity is weak for long wavelength waves, but becomes strong for short wavelength waves. At these scales  weak wave turbulence is no longer applicable. There is reason to believe that at very short scales the spectrum is again scale-invariant, but with a different exponent, referred to as the generalized Phillips spectrum  \cite{ Newell}. There is, however,  limited analytic understanding of strong wave turbulence. 

Wave turbulence, unlike the perhaps more familiar turbulence associated with the Navier-Stokes equation and the corresponding nonperturbative transfer of energy between eddies across scales, is a general phenomenon in nonlinear physics. For any Hamiltonian with small nonlinearity, exciting waves in some range of wavenumbers may trigger a cascade of energy across scales. And, unlike the case of Navier-Stokes turbulence, the ability to take a Hamiltonian of our choosing and to go to a regime with weak interactions, makes wave turbulence amenable to a systematic analytic treatment
\cite{RS1, RS2, RSSS, FR}.

There is a long history in quantum field theory and statistical physics of studying theories with a large number of fields ($N \gg 1$)\cite{Coleman, Wilson:1973jj, PhysRevD.10.3235,Klebanov:2018fzb}, and that is what we will do here. 
A physical example one can keep in mind is that of spin waves, in which the spins are in vector representations of $O(N)$.  We obtain the large $N$ wave kinetic equation, which, unlike the standard (weak) wave kinetic equation, is valid at arbitrarily strong nonlinearity (the case of a nonlinear interaction that is momentum independent was studied in \cite{Walz:2017ffj,bergesGasenzerScheppach2010, berges2002,bergesRothkopfSchmidt2008}). However, finding stationary solutions of this equation is challenging, as this is an integral equation. We therefore introduce a second major simplifying feature: we take the interactions to be concentrated about nearly equal momenta \cite{DYACHENKO, Hasselmann85, hasselmann1985computations, zakharov1999diffusion,npg-9-355-2002}.~\footnote{Something similar occurs  in the context of particle scattering, in which the Boltzmann equation becomes a differential equation for small-angle grazing collisions in plasmas, described by the Fokker-Planck/Landau kinetic equation \cite{Liboff, PhysRev.162.186,PhysRev.107.1}.} We refer to this as strongly local interactions, where  locality is in wavenumber space. This choice of interaction turns the integral equation into a differential equation, which is then straightforward to study and allows us to construct stationary solutions valid at all scales. 

 In Sec.~\ref{sec2} we present the kinetic equation in the strongly local and large $N$ limit. In Sec.~\ref{sec3} we show that asymptotically, at large or small wavenumber, there are three possible scaling solutions: the Kolmogorov-Zakharov (KZ) solution at weak nonlinearity, and either the Phillips solution, or what we refer to as the strong turbulence solution, at strong nonlinearity. In Sec.~\ref{sec4} we find stationary solutions of the kinetic equation at all scales which interpolate between these three asymptotic scalings. We conclude in Sec.~\ref{sec5}.

  \section{Large $N$ and strongly local kinetic equation}  \label{sec2}
Waves interacting with a quartic interaction have the Hamiltonian, 
  \be \label{H1}
H= \sum_p \o_p a^*_p a_p + \sum_{p_1,\ldots, p_4} \lam_{p_1 p_2 p_3 p_4} a^*_{p_1}a^*_{p_2} a_{p_3} a_{p_4}~,
\ee
where  $a_k$ is a complex scalar field,  $\o_k$ is the dispersion relation, and $\lam_{p_1p_2p_3p_4}$ is the interaction. For instance, for the nonlinear Schr\"odinger (Gross-Pitaevskii) equation  $\o_k\sim k^2$ and $\lam_{1234}$ is a constant. For surface gravity waves, after a canonical transformation to eliminate the nonresonant cubic interaction, the leading order term in the Hamiltonian takes the form (\ref{H1}) with $\o_k\sim \sqrt{k}$ and $\lam_{p_1p_2p_3p_4}$  a nontrivial scale-invariant function with scaling dimension three. For spin waves, the interaction is $ \lam_{p_1 p_2 p_3 p_4}\sim \v p_1 {\cdot} \v p_2 + \v p_3 {\cdot} \v p_4$. 

 Using this Hamiltonian, it is straightforward to show that weakly interacting waves are described by the kinetic equation \cite{Falkovich, Nazarenko}
\be \label{WKE}
\frac{\d n_k}{\d t} = 4\pi\!\!\! \sum_{p_1,\ldots, p_4} (\delta_{k p_1}{+}\delta_{kp_2}{-}\delta_{k p_3}{-}\delta_{kp_4}) |\lam_{p_1 p_2p_3p_4}|^2  \prod_{i=1}^4 n_i\, \Big( \frac{1}{n_1} {+} \frac{1}{n_2}{-}\frac{1}{n_3} {-} \frac{1}{n_4} \Big)\delta(\o_{p_1}{+}\o_{p_2}{-} \o_{p_3}{-}\o_{ p_4})~,
\ee
where $\delta_{kp_1}$ is the Kronecker delta function, $n_k$ is the occupation number, $n_k = \la a_k^* a_k\ra$, and this equation is valid to leading order in the interaction. 
This is the wave analog of the Boltzmann equation, which governs a dilute gas of particles.

The kinetic equation is an integral equation, and challenging to study in general. It drastically simplifies if one takes strongly local interactions: $\lam_{p_1 p_2 p_3 p_4}$ which is strongly peaked around momenta that are nearly equal $\v p_1 \approx \v p_2 \approx \v p_3 \approx \v p_4$. The kinetic equation then becomes a differential equation \cite{DYACHENKO}, 
 \be \label{diffKE}
\o^{\frac{d-\al}{\al}} \frac{\d n}{\d t} = \mathbb{\lam}^2 \frac{\d^2}{\d \o ^2}\( \o^{\frac{2\beta + 3d}{\al} +2}n^4 \frac{\d^2}{\d \o^2} \frac{1}{n}\)~,
 \ee
 where $n$ is a function of frequency $\o$ and time $t$, $d$ is the spatial dimension, $\lam^2$ is a dimensionful and system specific constant (which involves integrating the square of the interaction), and $\o_k$ and $\lam_{p_1 p_2 p_3 p_4}$ are assumed to be scale invariant functions: $\o_k\sim k^{\al}$ and $\lam_{p_1 p_2 p_3 p_4}\sim p^{\beta}$. This differential approximation to the kinetic equation is simple to see: the difference of the $1/n_i$ terms in (\ref{WKE}) vanishes at leading order if all the momenta are set equal. One must therefore Taylor expand, to second order, which gives the factor of $\frac{\d^2}{\d\o^2}\frac{1}{n}$. Likewise, the difference of the Kronecker delta functions in (\ref{WKE}) vanishes at leading order, and so Taylor expanding gives a second derivative.

 As mentioned earlier, the kinetic equation (\ref{WKE}) is only valid at leading order in the interaction. In particular, the only process it captures is one in which waves of momenta $p_1$ and $p_2$ scatter directly into $p_3$ and $p_4$. To go to higher order, one needs to include processes with multiple scatterings. In general, the number of processes and the number of intermediate states is enormous, growing exponentially with the order in $\lam$. To have any hope of studying strongly interacting waves, one must introduce some additional small parameter into the theory that will give preference to a subclass of processes. 
 
A common technique in quantum field theory and statistical physics is to consider a large $N$ theory \cite{Coleman, Wilson:1973jj, PhysRevD.10.3235,Klebanov:2018fzb}. Instead of one field, there are $N$ fields $a_p^j$, $i=1,\ldots, N$, which are grouped into a vector $\vec a_p$. The Hamiltonian is then, 
\be \label{HN}
H=  \sum_p \o_p \, {\vec a}^{\, *}_p{\cdot} \vec a_p + \frac{1}{N}\sum_{p_1, \ldots, p_4} \lam_{p_1 p_2 p_3 p_4} ({\vec a}^{\, *}_{p_1}{\cdot} \vec a_{p_3})({\vec a}^{\, *}_{p_2} {\cdot} \vec a_{p_4})~.
\ee
Introducing more fields may seem like a complication, rather than a simplification. The key, however, is to take $N$ to be large while keeping $\lam_{p_1 p_2 p_3 p_4}$ finite. This is the large $N$ limit.

%%%%%%
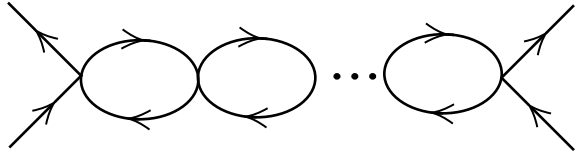
\begin{figure}[t]
\centering
\tikzset{every picture/.style={line width=1pt}} %set default line width to 0.75pt        
\begin{tikzpicture}[x=0.75pt,y=0.75pt,yscale=-1,xscale=1]
%Straight Lines [id:da041973723771052684] 
\draw    (120.99,79.59) -- (99.61,58.21) ;
\draw [shift={(98.2,56.8)}, rotate = 45] [color={rgb, 255:red, 0; green, 0; blue, 0 }  ][line width=0.75]    (10.93,-4.9) .. controls (6.95,-2.3) and (3.31,-0.67) .. (0,0) .. controls (3.31,0.67) and (6.95,2.3) .. (10.93,4.9)   ;
%Straight Lines [id:da6821573218610524] 
\draw    (84.2,42.8) -- (98.2,56.8) ;

%Straight Lines [id:da9765176178639232] 
\draw    (84.87,115.92) -- (106.24,94.55) ;
\draw [shift={(107.65,93.13)}, rotate = 135] [color={rgb, 255:red, 0; green, 0; blue, 0 }  ][line width=0.75]    (10.93,-4.9) .. controls (6.95,-2.3) and (3.31,-0.67) .. (0,0) .. controls (3.31,0.67) and (6.95,2.3) .. (10.93,4.9)   ;
%Straight Lines [id:da5458905743847126] 
\draw    (121.65,79.13) -- (107.65,93.13) ;

%Straight Lines [id:da5198985984190069] 
\draw    (369.65,117.25) -- (348.28,95.88) ;
\draw [shift={(346.87,94.47)}, rotate = 45] [color={rgb, 255:red, 0; green, 0; blue, 0 }  ][line width=0.75]    (10.93,-4.9) .. controls (6.95,-2.3) and (3.31,-0.67) .. (0,0) .. controls (3.31,0.67) and (6.95,2.3) .. (10.93,4.9)   ;
%Straight Lines [id:da08815724901864475] 
\draw    (332.87,80.47) -- (346.87,94.47) ;

%Shape: Ellipse [id:dp053206746908869285] 
\draw   (120.99,81.71) .. controls (120.99,70.69) and (134.24,61.75) .. (150.59,61.75) .. controls (166.94,61.75) and (180.2,70.69) .. (180.2,81.71) .. controls (180.2,92.73) and (166.94,101.67) .. (150.59,101.67) .. controls (134.24,101.67) and (120.99,92.73) .. (120.99,81.71) -- cycle ;
%Straight Lines [id:da2049255397226224] 
\draw    (156.11,62.11) -- (152.59,61.88) ;
\draw [shift={(152.59,61.88)}, rotate = 183.71] [color={rgb, 255:red, 0; green, 0; blue, 0 }  ][line width=0.75]    (10.93,-4.9) .. controls (6.95,-2.3) and (3.31,-0.67) .. (0,0) .. controls (3.31,0.67) and (6.95,2.3) .. (10.93,4.9)   ;
%Straight Lines [id:da8364971546932667] 
\draw    (150.59,101.67) -- (146.93,101.45) ;
\draw [shift={(144.94,101.33)}, rotate = 3.37] [color={rgb, 255:red, 0; green, 0; blue, 0 }  ][line width=0.75]    (10.93,-4.9) .. controls (6.95,-2.3) and (3.31,-0.67) .. (0,0) .. controls (3.31,0.67) and (6.95,2.3) .. (10.93,4.9)   ;
%Straight Lines [id:da011057961325504961] 
\draw    (332.87,80.92) -- (354.24,59.55) ;
\draw [shift={(355.65,58.13)}, rotate = 135] [color={rgb, 255:red, 0; green, 0; blue, 0 }  ][line width=0.75]    (10.93,-4.9) .. controls (6.95,-2.3) and (3.31,-0.67) .. (0,0) .. controls (3.31,0.67) and (6.95,2.3) .. (10.93,4.9)   ;
%Straight Lines [id:da7710982612102945] 
\draw    (369.65,44.13) -- (355.65,58.13) ;

%Shape: Ellipse [id:dp6413775230121093] 
\draw   (179.99,80.71) .. controls (179.99,69.69) and (193.24,60.75) .. (209.59,60.75) .. controls (225.94,60.75) and (239.2,69.69) .. (239.2,80.71) .. controls (239.2,91.73) and (225.94,100.67) .. (209.59,100.67) .. controls (193.24,100.67) and (179.99,91.73) .. (179.99,80.71) -- cycle ;
%Straight Lines [id:da8305176459189529] 
\draw    (215.11,61.11) -- (211.59,60.88) ;
\draw [shift={(211.59,60.88)}, rotate = 183.71] [color={rgb, 255:red, 0; green, 0; blue, 0 }  ][line width=0.75]    (10.93,-4.9) .. controls (6.95,-2.3) and (3.31,-0.67) .. (0,0) .. controls (3.31,0.67) and (6.95,2.3) .. (10.93,4.9)   ;
%Straight Lines [id:da46676391799209416] 
\draw    (209.59,100.67) -- (205.93,100.45) ;
\draw [shift={(203.94,100.33)}, rotate = 3.37] [color={rgb, 255:red, 0; green, 0; blue, 0 }  ][line width=0.75]    (10.93,-4.9) .. controls (6.95,-2.3) and (3.31,-0.67) .. (0,0) .. controls (3.31,0.67) and (6.95,2.3) .. (10.93,4.9)   ;
%Shape: Ellipse [id:dp9640096501828748] 
\draw   (273.99,78.71) .. controls (273.99,67.69) and (287.24,58.75) .. (303.59,58.75) .. controls (319.94,58.75) and (333.2,67.69) .. (333.2,78.71) .. controls (333.2,89.73) and (319.94,98.67) .. (303.59,98.67) .. controls (287.24,98.67) and (273.99,89.73) .. (273.99,78.71) -- cycle ;
%Straight Lines [id:da7487756183499301] 
\draw    (309.11,59.11) -- (305.59,58.88) ;
\draw [shift={(305.59,58.88)}, rotate = 183.71] [color={rgb, 255:red, 0; green, 0; blue, 0 }  ][line width=0.75]    (10.93,-4.9) .. controls (6.95,-2.3) and (3.31,-0.67) .. (0,0) .. controls (3.31,0.67) and (6.95,2.3) .. (10.93,4.9)   ;
%Straight Lines [id:da8473885495729663] 
\draw    (303.59,98.67) -- (299.93,98.45) ;
\draw [shift={(297.94,98.33)}, rotate = 3.37] [color={rgb, 255:red, 0; green, 0; blue, 0 }  ][line width=0.75]    (10.93,-4.9) .. controls (6.95,-2.3) and (3.31,-0.67) .. (0,0) .. controls (3.31,0.67) and (6.95,2.3) .. (10.93,4.9)   ;
%Shape: Ellipse [id:dp5027967650560022] 
\draw  [fill={rgb, 255:red, 0; green, 0; blue, 0 }  ,fill opacity=1 ] (249,80) .. controls (249,79.45) and (249.5,79) .. (250.11,79) .. controls (250.72,79) and (251.22,79.45) .. (251.22,80) .. controls (251.22,80.55) and (250.72,81) .. (250.11,81) .. controls (249.5,81) and (249,80.55) .. (249,80) -- cycle ;
%Shape: Ellipse [id:dp2675779869113736] 
\draw  [fill={rgb, 255:red, 0; green, 0; blue, 0 }  ,fill opacity=1 ] (258,80) .. controls (258,79.45) and (258.5,79) .. (259.11,79) .. controls (259.72,79) and (260.22,79.45) .. (260.22,80) .. controls (260.22,80.55) and (259.72,81) .. (259.11,81) .. controls (258.5,81) and (258,80.55) .. (258,80) -- cycle ;
%Shape: Ellipse [id:dp791991204775183] 
\draw  [fill={rgb, 255:red, 0; green, 0; blue, 0 }  ,fill opacity=1 ] (267,80) .. controls (267,79.45) and (267.5,79) .. (268.11,79) .. controls (268.72,79) and (269.22,79.45) .. (269.22,80) .. controls (269.22,80.55) and (268.72,81) .. (268.11,81) .. controls (267.5,81) and (267,80.55) .. (267,80) -- cycle ;
\end{tikzpicture} 
\caption{With a large $N$ number of fields, rather than  one field (\ref{H1}), the  process of two wave scattering is dominated by bubble diagrams. These can be summed, allowing one to study the theory at strong nonlinearity.}\label{bubblesum}
\end{figure}
%%%%%%%%%%%%

At leading nontrivial order in $1/N$ the only processes that appear for the scattering of two waves are the ``bubble diagrams'', shown in Fig.~\ref{bubblesum}, see Appendix~\ref{apA}. 
Each bubble gives an additional factor proportional to $\lam \o^{\frac{d+\beta}{\al}} \frac{\d n}{\d \o}$. Summing the geometric series of bubble diagrams gives the kinetic equation,
\be \label{KE}
\o^{\frac{d-\al}{\al}} \frac{\d n}{\d t} =\frac{1}{N}\frac{\partial^2}{\partial\omega^2}\(\frac{\lam^2 \omega^{\frac{2\beta+3d}{\alpha}+2}n^4\frac{\partial^2}{\partial\omega^2}\frac{1}{n}}{\left|1- c\, \lam\, \omega^\frac{d+\beta}{\alpha}\der{n}{\omega}\right|^2}\),\
\ee
where $c$ is a system specific constant (and generally complex.  For simplicity, in what follows we treat $c$ as real; the more general case is done in Appendix~\ref{apD}).  This equation is valid at all $\lam$ and at all $\o$. 
Notice that in the limit of small $\lam$, we may drop the $\lam$ term in the denominator and  recover the weak kinetic equation (\ref{diffKE}). 

The stationary solutions correspond to setting the left hand side of (\ref{KE}) to zero. The right hand side must therefore be of the form of a second derivative of $P-Q \omega$, where $P$ is the energy flux and $Q$ is the particle number flux, as can be seen by noticing that (\ref{KE}) is of the form of a continuity equation, $\frac{\d n}{\d t} =- \frac{\d J}{\d \o}$. The energy flux $P$ should be positive (direct cascade) whereas the number flux $Q$ should be negative (inverse cascade).  Rearranging  gives,
\be
	\lam^2 \omega^2	\frac{\partial^2}{\partial\omega^2}\frac{1}{n}={\left(P{-}Q \omega\right)\omega^{-\frac{3d+2\beta}{\alpha}}}\left(n^{-2} -c\, \lam\, \omega^\frac{d+\beta}{\alpha}\der{}{\omega}\frac{1}{n}\right)^2~.\label{26}
\ee
Our goal is to find solutions of this equation.

 \section{Three asymptotic solutions}   \label{sec3}
 There are three terms in (\ref{26}). We may easily find a power law solution to the equation if we drop any one of these three terms. We consider all three options.  
 
 \subsection*{Kolmogorov-Zakharov}
 Dropping the term proportional to $\lam$ should take us back to the weak wave turbulence (Kolmogorov-Zakharov) solution. Indeed, doing this gives, 
 \be\label{eqAsympWeak}
 \lam^2 \o^{\frac{2\beta + 3d}{\al} +2}n^4 \frac{\d^2}{\d \o^2} \frac{1}{n} = P-Q\o~.
 \ee
 Setting $Q=0$ (constant energy flux), and inserting $n\sim k^{-\g} \sim \o^{-\g/\al}$ gives the Kolmogorov-Zakharov (KZ) solution for an energy cascade, while  setting $P=0$ (constant number flux), gives the KZ solution for a particle number cascade:
 \bea \label{KZ1}
 n&\sim& \lambda^{-2/3}P^{1/3}\o^{-\g/\al}~, \ \ \ \  \ \ \  \ \  \g = d + \frac{2}{3}\beta~, \ \ \ \  \ \  \ \ \ \ Q=0\\ \label{KZ2}
  n &\sim& \lambda^{-2/3} (-Q)^{1/3}\o^{-\g/\al}~, \ \ \ \ \g = d + \frac{2}{3}\beta - \frac{\al}{3}~, \ \ \ \ P=0~.
  \eea
  Setting both $P$ and $Q$ to zero gives the thermal solution, $n \sim \frac{1}{\o + \mu}$ where $\mu$ is a constant (chemical potential).

  \subsection*{Strong turbulence}
  We now instead drop the $n^{-2}$ term in (\ref{26}), corresponding to large $\lam$, and leaving us with, 
\be \label{eqAsympStrong}
	\omega^2	\frac{\partial^2}{\partial\omega^2}\frac{1}{n}={\left(P-Q \omega\right)\omega^{-\frac{3d+2\beta}{\alpha}}}\left(c\, \omega^\frac{d+\beta}{\alpha}\der{}{\omega}\frac{1}{n}\right)^2~.
\ee
Depending on if we set $Q$ or $P$ to zero, we obtain power law solutions \cite{bergesGasenzerScheppach2010}, 
 \bea \label{S1}
 n&\sim& c^2 P\o^{-\g/\al}~, \ \ \ \ \ \ \ \g = d + 2\al~, \ \ \ \  \ \  \ \ \ \ Q=0\\
  n &\sim&  -c^2 Q \o^{-\g/\al}~, \ \ \ \ \ \g = d +\al~, \ \ \ \ \ \ \ \ \ \ \  P=0~.
  \eea
  
   \subsection*{Phillips (Critical Balance)}
   The final possibility is to drop the left-hand side of (\ref{26}). The limit in which this is justified  is large $\lam$ with $n\sim 1/\lam$, so that the left-hand side is of order $\lam^3$, whereas the right-hand side is of order $\lam^4$, and therefore dominant. Note that this is only possible if $c\, \lam$ is real and positive. Dropping the left-hand side  leaves us with, 
   \be\label{eqAsympPhillips}
   n^{-2} -c\, \lam\, \omega^\frac{d+\beta}{\alpha}\der{}{\omega}\frac{1}{n}=0~,
   \ee
   and hence, 
   \be \label{38}
   n \sim \frac{1}{c\, \lam} \o^{\frac{-\g}{\al}}~, \ \ \ \ \g = d+\beta-\al~.
   \ee
   This is known as the generalized Phillips  \cite{Newell, Phillips}, or critical balance, solution \cite{Goldreich}.~\footnote{The possibility of obtaining the Phillips solution from summing bubble diagrams was suggested in \cite{FR}.}\\
   
   The strength of the nonlinearity is commonly parametrized by $\eps_k$, the ratio of the quartic to quadratic terms in the Hamiltonian in a state with occupation numbers $n_k$, 
   \be \label{defEpsilon}
   \eps_k = \frac{\lam_{k k k k} n_k k^d }{\o_k}\sim k^{\beta+ d- \gamma - \al}~.
   \ee    
For us, this is conveniently just the ratio of the second to the first term on the right side of (\ref{26}). The Phillips solution (\ref{38}) corresponds to $\eps_k$ that is constant ($k$ independent). 
   On the other hand, for the KZ direct cascade $\eps_k\sim k^{\beta/3 - \al}$ for the energy flux solution (\ref{KZ1}), which  means that we expect KZ to be valid at small $k$ if $\beta>3\al$. At large $k$ we expect to generically have the strong turbulence scaling (\ref{S1}) or potentially the Phillips scaling (\ref{38}). For $\beta< 3\al$ the situation is reversed: we expect KZ scaling at large $k$ and strong turbulence or Phillips scaling at small $k$. We must solve the full differential equation (\ref{26}) to see explicitly how the transition occurs. This is what we do next.

\section{Turbulent solutions at all scales}  \label{sec4}

Our task now is to find solutions of the differential equation (\ref{26}) governing stationary solutions of the  strongly local large $N$ model. 

\subsection*{Energy cascade}
Let us focus on energy cascades ($Q=0$) for now. It is convenient to change variables,
\be \label{41}
n=\(\frac{P}{\lam^2}\)^{1/3}\frac{1}{\o^\zeta}\frac{1}{u}~, \ \ \zeta= \frac{d + \frac{2}{3}\beta}{\al}~,\  \ \ \ \tau= \log\frac{\o}{\o_0}~, 
\ee
where $\o_0$ is an arbitrary scale. 
 This transforms (\ref{26}) into, 
\be \label{diffeq}
\ddot{u} + (2\zeta{-}1) \dot{u} + \zeta(\zeta{-}1) u = \(u^2 -  \eps (\zeta u + \dot u)\)^2~, \ \ \ \ \ \eps \equiv  c\, \left(\lam P \omega^{\frac{\beta- 3\al}{\alpha}}\right)^{1/3}~,
\ee 
where a dot denotes a derivative with respect to $\tau$ (which should not be confused with physical time $t$ in (\ref{KE})). The limit of small $\eps $  reduces this to the weak coupling equation, and the stationary solution is found by simply setting $\ddot u = \dot u=0$ and hence $\zeta(\zeta{-}1)u = u^4$, which is just the KZ solution (\ref{KZ1}). 

As long as $\beta \neq 3\al$, the nonlinearity parameter $\eps$ is $\tau$ dependent, and so regardless of the interaction strength $\lam$ and of the flux $P$, at either large or small $\tau$, $\eps$ will become large, i.e.,  $\eps$ is invariant under a shift of $\tau \rightarrow \tau+ \tau_0$ and a rescaling $\lam P\rightarrow \lam Pe^{-\frac{\beta-3\alpha}{\alpha}\tau_0}$. 
The simplest case we may consider is for a constant $\eps$, which occurs for $\beta = 3\al$. The differential equation (\ref{diffeq}) then has no explicit $\tau$ dependence, and finding the stationary solutions reduces to solving an algebraic cubic equation, $\zeta(\zeta{-}1) = u\(u {-}  \eps\, \zeta\)^2$. In  the small $\eps$ limit only one of the solutions is real, and is identified as the Kolmogorov-Zakharov solution. In the large $\epsilon$ limit there are three real solutions,  two of which are the Phillips solution and one of which is the strong turbulence solution (if $c^3 \lam$, and correspondingly $\eps$, is positive. Otherwise,  the only real solution is the strong turbulence one).
Now consider the case of general $\beta$. The differential equation (\ref{diffeq}) now has explicit $\tau$ dependence. Perturbatively expanding around the three solutions in the previous section, to next-to-leading order, and defining $\kappa =( \frac{\beta}{3} - \al)/\al$ gives,  
 \bea \nn
 u &=& (\zeta(\zeta{-}1))^{\frac{1}{3}} + \frac{2 \zeta^2 (\zeta {-}1)}{3\zeta (\zeta{-}1)-  (2\zeta{-}1)\kappa -\kappa^2} \eps + \ldots~, \ \ \ \  \ \ \ \  \  \eps\ll1~, \ \ \ \text{Kolmogorov-Zakharov}~, \\ \nn
 u &=& \frac{\zeta{-}1{-}2\kappa}{\zeta{-}2\kappa} \frac{1}{\eps^2} + \frac{2(\zeta{-}1{-}2\kappa)^3}{(\zeta{-}2\kappa)^2(\zeta{-}5\kappa)(\zeta{-}1{+}\kappa)}\frac{1}{\eps^5} + \ldots~, \ \ \ \ \ \ \eps\gg 1~, \ \ \ \text{strong turbulence}~,\\ 
 u &=& (\zeta{+} \kappa) \, \eps \pm\frac{\zeta{+}\kappa}{\zeta{+}\frac{5}{2}\kappa}\frac{\sqrt{\zeta{-}1 {+}\kappa}}{\sqrt{\eps}}+ \ldots~, \ \ \ \ \ \  \ \ \ \  \ \ \  \  \ \ \ \  \ \ \ \  \ \  \ \ \eps\gg 1~, \  \  \ \text{Phillips}~, \label{47}
  \eea
  where we used that the $\tau$ dependence is within $\eps$, i.e., $\dot{\eps} =\kappa \eps$.

\begin{figure}[]
\subfloat[]{ \includegraphics[width=0.45\textwidth]{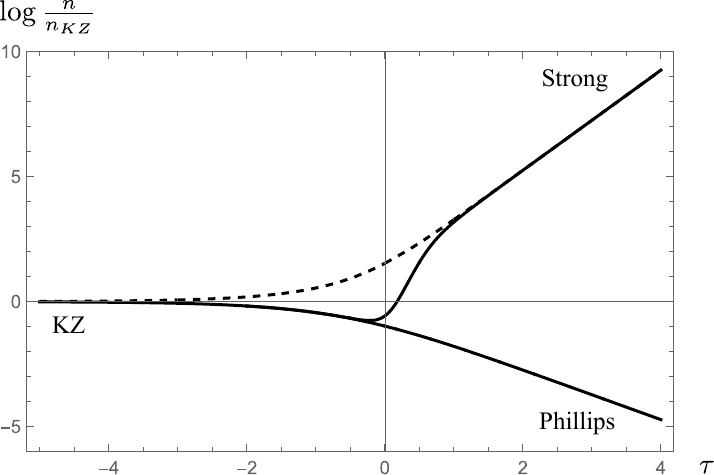}} \ \ \ \ \ \ \ \ \ \ \ 
	\subfloat[]{\includegraphics[width=0.45\textwidth]{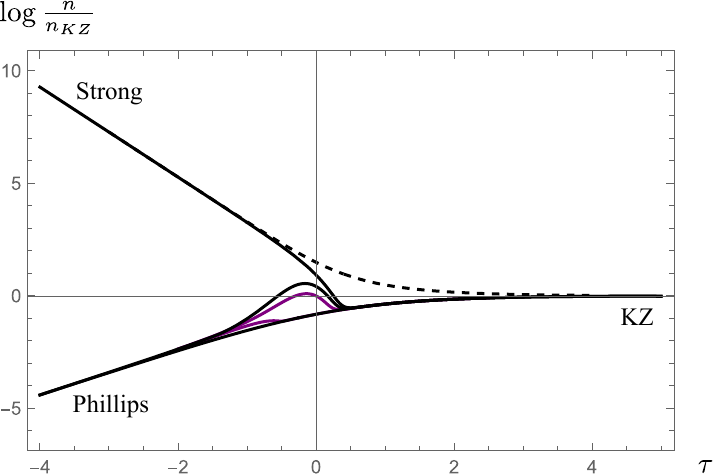} }
		\centering
		\caption{Stationary solutions of the kinetic equation (\ref{KE}) for the strongly local large $N$ model. Log-log plot of $n$ vs $\omega=\omega_0 e^\tau$. $n$ is normalized by a pure Kolmogorov-Zakharov (KZ) solution $n_{KZ}$ defined in \eqref{KZ1}. The parameters are taken to be $d=2, \alpha{=}1/2$ and  $\beta=3$, $\eps(\tau{=}0)^3 = 2/3$ in (a) and $\beta=0$, $\eps(\tau{=}0)^3 = 3/2$ in (b).  The solutions interpolating between KZ and either strong or Phillips behavior are shown with solid lines. In (b) there is a continuum of solutions interpolating between Phillips and KZ (some examples shown in purple), lying between two boundary curves (black). The dotted lines are the KZ to strong solutions for the case where $\lambda c$ is real and negative.}\label{fig1}
\end{figure}

To see the full solution for all $\omega$ we must numerically solve the differential equation (\ref{diffeq}); see Appendix~\ref{appendixB} for details. 
There are two qualitatively different cases, depending on if $\beta/3$ is greater than or less than $\al$, which determines if Kolmogorov-Zakharov scaling occurs at small $\o$ or at large $\o$. We solve (\ref{diffeq}) for a representative example of each of these two cases, see Fig.~\ref{fig1}. We find solutions that interpolate between Kolmogorov-Zakharov scaling at one end (large $\o$ for $\beta/3{<}\al$ and small $\o$ for $\beta/3{>}\al$) and strong turbulence scaling (\ref{S1}) or Phillips scaling (\ref{38}) on the other end. Phillips scaling can only be realized for $c \lam$ that is positive. Which scaling the solution asymptotes to, Phillips or strong turbulence, depends on the boundary conditions.  

Stationary solutions for the differential model of weak wave turbulence (the differential equation (\ref{diffeq}) with $\eps = 0$) were studied in \cite{Naz}. The generic solution  is a combination of KZ scaling in some range of $\omega$, and then either thermal scaling at the ends, or a fall off behavior ($n \sim (\o - \o_*)^{2/3}$ is a solution in the vicinity of some arbitrary $\o_*$). Likewise, in the case of nonzero $\eps$ one can have solutions that follow the black lines in Fig.~\ref{fig1}  for some range of $\o$ and then  either fall off or asymptote to thermal behavior.

\subsection*{Dual cascade}

\begin{figure}[]
\subfloat[]{
	\includegraphics[width=0.45\textwidth]{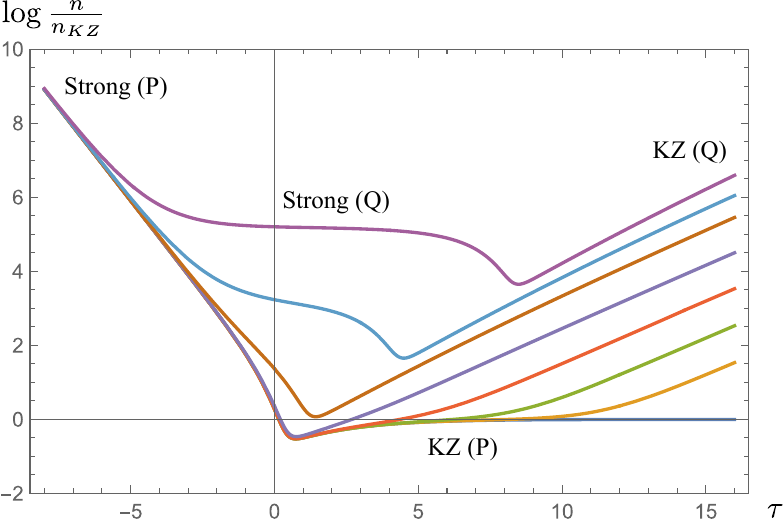}} \ \ \ \ \ \ \ \ \ \ \ 
	\subfloat[]{
	\includegraphics[width=0.45\textwidth]{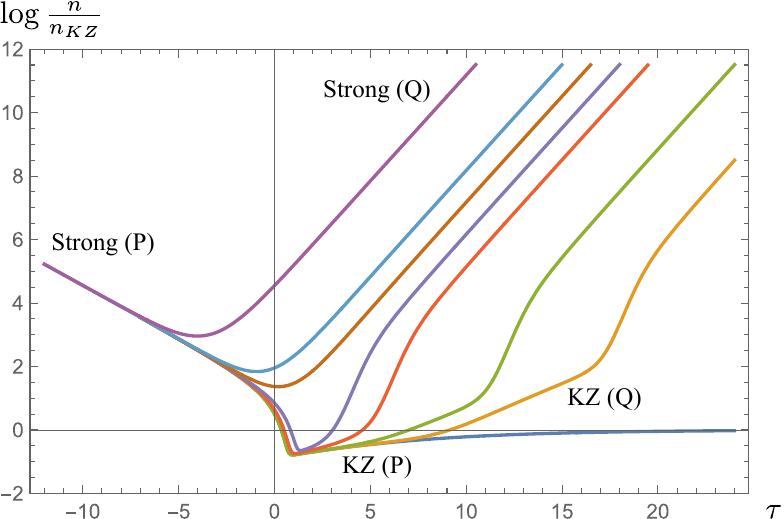}}
	\centering
	\caption{Log-log plots of $n$ vs $\omega$ for the dual cascade case. a) $d{=}2, \alpha{=}2, \beta{=}3$. The dimensionless quantity $\tilde{P}=c^3\omega_0^{3\kappa}\lambda P$ is held fixed at $3/2$ for all curves and $Q$ is varied. From top to bottom the values of $\log\left(-\frac{Q}{P}\omega_0\right)$ for the curves are are $4, 2, 0, -3, -6, -9, -12, -\infty$. b) Values of the parameters are chosen such that there is strong coupling in both the IR and UV. $d=2, \alpha=2, \beta=5, \tilde{P}=81/50$. From top to bottom the values of $\log\left(-\frac{Q}{P}\omega_0\right)$ for the curves are $3, 0, -1, -2, -3, -6, -9, -\infty$.}\label{figDual}
\end{figure}

So far we have considered  energy cascades. The analysis for a particle number cascade is similar. An even richer case is a dual cascade of both nonzero energy flux $P$ and particle number flux $Q$. In this case the differential equation (\ref{26}) has unavoidable explicit $\omega$ dependence, even in the weak coupling limit.
 It is convenient to use slightly different variables, $v$ instead of $u$, defined as
\be \label{44}
n=\frac{1}{c\, \lam}\frac{1}{\o^{\xi}}\frac{1}{v}~, \ \ \xi = \frac{d + \beta-\alpha}{\al}~,
\ee
which transforms (\ref{26}) into, 
\begin{align}\label{eqMain}
\ddot{v} + (2\xi{-}1) \dot{v} +\xi(\xi{-}1) v = \epsilon^3\(v^2 -(\xi v + \dot v)\)^2~, \ \ \ \ \ \epsilon^3\equiv c^3 \lam(P-Q \omega )\omega^{\frac{\beta-3\alpha}{\alpha}}.
\end{align}
An example of a dual cascade is shown in Fig.~\ref{figDual}(a). All solutions interpolate between strong energy flux scaling for small $\omega$ and  KZ number flux scaling for large $\omega$, with intermediate ranges of either strong number flux  or KZ energy flux scaling (if $Q\omega_0$ is much larger than or much smaller than $P$, respectively).

\begin{figure}[]
\subfloat[]{	\includegraphics[width=0.45\textwidth]{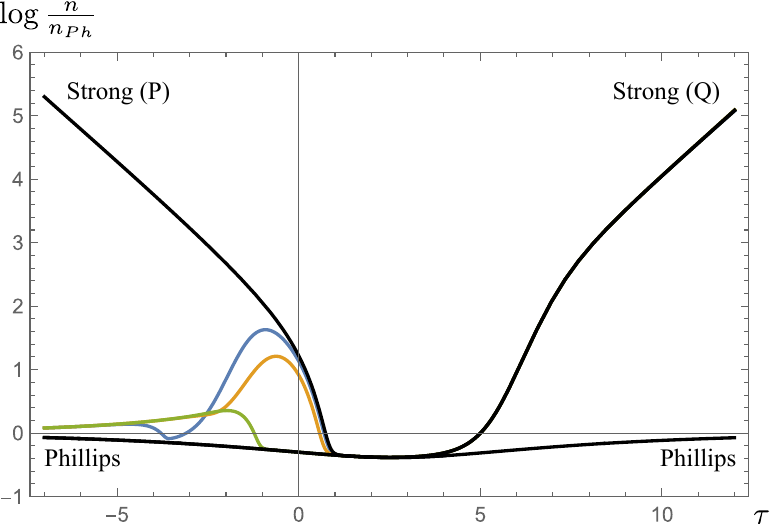}} \ \ \ \ \ \ \ \ \ \ \  
\subfloat[]{
\includegraphics[width=0.45\textwidth]{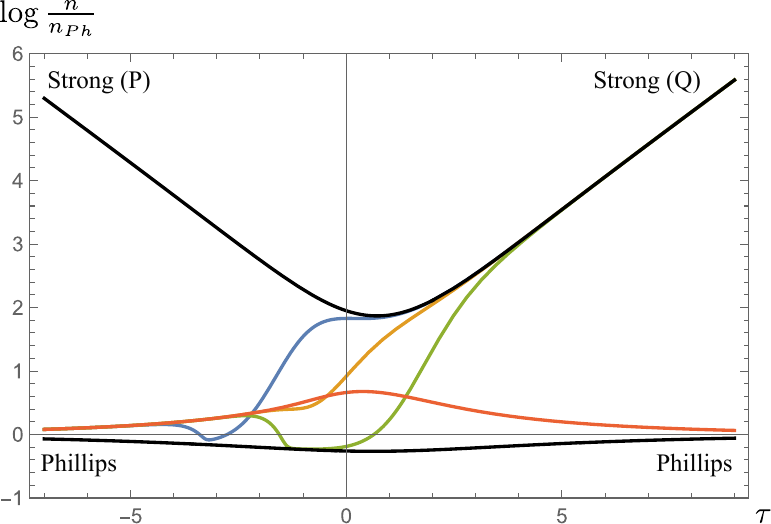}}	\centering
	\caption{Solutions interpolating between strong turbulence and Phillips in the dual cascade. The vertical axis is normalized by the pure Phillips solution  $n_{Ph}$ in \eqref{38} and the parameters are identical to Fig~\ref{figDual}(b). a) $\log\left(-\frac{Q}{P}\omega_0\right)=-3$. All four combinations involving Phillips or strong turbulence in the IR and UV are possible (black). As in Fig~\ref{fig1}(b), there is an additional family of solutions with Philips in the IR (e.g. blue, orange, green). b) $\log\left(-\frac{Q}{P}\omega_0\right)=-1$. Solutions interpolating between strong turbulence in the IR and Phillips in the UV are no longer possible. Phillips in the IR interpolating to strong turbulence in the UV is still possible (e.g. blue, orange, green) but disappears for even higher flux. }\label{figDualPhillips}
\end{figure}

A particularly interesting dual cascade occurs for $\beta$ that lie in the range $2\al<\beta<3\al$. In this case the nonlinearity parameter $\eps$ is large at  both ends, $\tau\rightarrow \pm \infty$, and so asymptotically there must be either strong or Phillips scaling, rather than Kolmogorov-Zakharov scaling. An example is shown in Fig.~\ref{figDual}(b). All solutions interpolate between strong energy flux scaling for small $\omega$ and strong number flux scaling for large $\omega$. For $Q\omega_0\ll P$ there may be intermediate regions with both energy flux and number flux KZ scaling. Fig.~\ref{figDual} concerns only strong turbulence, but there may also be Phillips scaling in the dual cascade, see Fig.~\ref{figDualPhillips}.


\subsection*{Phase portrait}
\begin{figure}[]
	\includegraphics[width=0.4\textwidth]{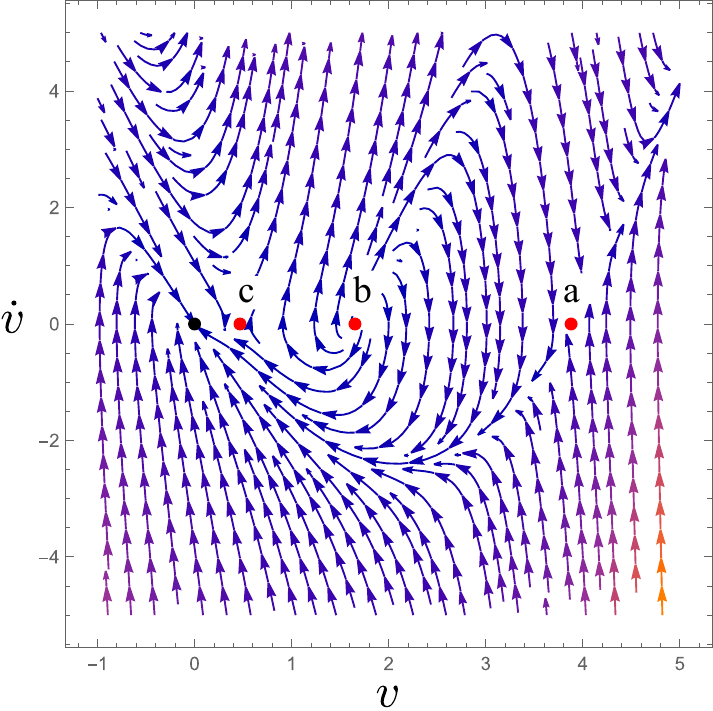}	\centering
	\caption{Phase portrait of (\ref{eqMain}), with $\xi =3$ and $\eps=2$,  treated as a dynamical system with variables $v, \dot v$.}\label{figPhasePlot}
\end{figure}

A phase portrait is sometimes a useful way of visualizing the solutions of a differential equation. Treating $v$ and $\dot v$ as independent variables, at each $(v, \dot v)$ one draws an arrow pointing in the direction of the $\tau$ derivative of $(v, \dot v)$.  If $\eps$ is a constant, a phase portrait captures all of the behavior. Since $\eps$ is generally $\tau$ dependent, one needs to know the  phase portrait at multiple values of $\eps$  to capture the qualitative behavior.

A phase portrait of the system at intermediate $\epsilon$ is shown in Fig.~\ref{figPhasePlot}, where the arrows point in the direction of increasing $\tau$. The running off to infinity of most trajectories at either small or large $\tau$ corresponds to $n$ either vanishing or diverging like a thermal spectrum with negative chemical potential \cite{Naz}. The trivial fixed point at $u=0$ is shown in black (and corresponds to thermal behavior), and the three non-trivial fixed points are shown in red and are labeled $a$, $b$, and $c$. The phase plot changes qualitatively upon varying the nonlinearity parameter $\epsilon$. As $\epsilon$ increases, we find that the fixed points $a$ and $b$ approach  the value corresponding to the Phillips scaling solution, while the fixed point $c$ corresponds to the strong scaling solution. In the opposite limit in which $\epsilon$ decreases, points $b$ and $c$ merge and disappear,  while $a$  corresponds to the  KZ solution.

We can explain various features of the numerical solutions by considering these fixed points. For instance, the curve in Fig.~\ref{fig1}(a)  going from KZ scaling to Phillips scaling corresponds to a path in phase space  remaining in the vicinity of point $a$ (which moves as $\tau$ changes). On the other hand, the curve going from KZ scaling to strong turbulence scaling involves  going from point $a$ to point $c$ in phase space, and there is a visible feature on the curve when this transition takes place. In the case of Fig.~\ref{fig1}(b), there are additional curves interpolating between Phillips and KZ scaling (purple) which may be understood as going from the IR attractive point $b$ at arbitrarily small $\omega$ to point $a$. In the dual cascade case of Fig.~\ref{figDualPhillips}, the value of $Q\omega_0/P$ determines some minimum value of $\epsilon$. In Fig.~\ref{figDualPhillips}(a) the minimum value is small enough such that the fixed point $c$ disappears for some intermediate value of $\tau$. As a result,  the strong energy flux to strong number flux solution must go from $c$ to $a$ and then back to $c$. In Fig.~\ref{figDualPhillips}(b) the minimum value of $\epsilon$ is large enough that the upper black strong turbulence curve can stay near the fixed point $c$ for all $\tau$. There is an additional Phillips solution (red) that stays near point $b$ for all $\tau$.

\section{Discussion}  \label{sec5}
Our main result is the kinetic equation (\ref{KE}). It goes beyond the standard kinetic theory for weak interactions, and is valid for arbitrarily strong interactions. We have found its stationary solutions, which realize the generalized Phillips spectrum (critical balance) and a new  strong wave turbulence spectrum. There is a simple argument for the Phillips spectrum, based on the hypothesis that the nonlinear term stops growing once it reaches the same order as the linear term \cite{Newell, Goldreich}. Likewise, the strong wave turbulence spectrum can be argued for on dimensional grounds, see Appendix~\ref{apC}. The challenge, however, is to have a consistent dynamical theory that achieves these scalings. Our kinetic equation (\ref{KE}) does just that. 

 It will be important to study time dependent solutions \cite{Connaughton_2003,Galtier:2000ce} of this kinetic equation inorder to understand the mechanism by which these turbulent cascades are dynamically formed. The strongly local interactions we took prevent any potential divergences (i.e., dependence on the pumping and dissipation scales), which can be physically relevant.  It will therefore be important to study wave turbulence at large $N$, with the assumption of strong locality relaxed. The large $N$ kinetic equation for general interactions is presented in  Appendix~\ref{apA}. 
 
\sss*{Acknowledgments} 
We thank G.~Falkovich  and M.~Smolkin for helpful discussions. 
This work  is supported in part by NSF grant PHY-2209116  and by the ITS through a Simons grant. 

\appendix
\section{ Large $N$ kinetic equation} \label{apA}

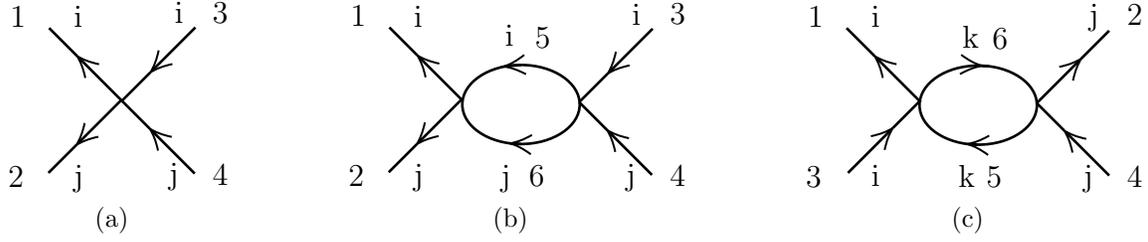
\begin{figure}[t]
\centering
\subfloat[]{
\tikzset{every picture/.style={line width=1pt}} %set default line width to 0.75pt        
\begin{tikzpicture}[x=0.75pt,y=0.75pt,yscale=-1,xscale=1]
%Straight Lines [id:da9153477132713148] 
\draw    (120.99,79.59) -- (99.61,58.21) ;
\draw [shift={(98.2,56.8)}, rotate = 45] [color={rgb, 255:red, 0; green, 0; blue, 0 }  ][line width=0.75]    (10.93,-4.9) .. controls (6.95,-2.3) and (3.31,-0.67) .. (0,0) .. controls (3.31,0.67) and (6.95,2.3) .. (10.93,4.9)   ;
%Straight Lines [id:da3850884700046392] 
\draw    (84.2,42.8) -- (98.2,56.8) ;

%Straight Lines [id:da8337941395021127] 
\draw    (120.65,79.13) -- (99.28,100.51) ;
\draw [shift={(97.87,101.92)}, rotate = 315] [color={rgb, 255:red, 0; green, 0; blue, 0 }  ][line width=0.75]    (10.93,-4.9) .. controls (6.95,-2.3) and (3.31,-0.67) .. (0,0) .. controls (3.31,0.67) and (6.95,2.3) .. (10.93,4.9)   ;
%Straight Lines [id:da11003978097052458] 
\draw    (83.87,115.92) -- (97.87,101.92) ;

%Straight Lines [id:da7117654134297163] 
\draw    (157.65,116.25) -- (136.28,94.88) ;
\draw [shift={(134.87,93.47)}, rotate = 45] [color={rgb, 255:red, 0; green, 0; blue, 0 }  ][line width=0.75]    (10.93,-4.9) .. controls (6.95,-2.3) and (3.31,-0.67) .. (0,0) .. controls (3.31,0.67) and (6.95,2.3) .. (10.93,4.9)   ;
%Straight Lines [id:da7664461985067011] 
\draw    (120.87,79.47) -- (134.87,93.47) ;

%Straight Lines [id:da4927483440171442] 
\draw    (157.99,42.47) -- (136.61,63.84) ;
\draw [shift={(135.2,65.25)}, rotate = 315] [color={rgb, 255:red, 0; green, 0; blue, 0 }  ][line width=0.75]    (10.93,-4.9) .. controls (6.95,-2.3) and (3.31,-0.67) .. (0,0) .. controls (3.31,0.67) and (6.95,2.3) .. (10.93,4.9)   ;
%Straight Lines [id:da7078994304202018] 
\draw    (121.2,79.25) -- (135.2,65.25) ;
% Text Node
\draw (63,30) node [anchor=north west][inner sep=0.75pt]   [align=left] {1};
% Text Node
\draw (62,111) node [anchor=north west][inner sep=0.75pt]   [align=left] {2};
% Text Node
\draw (165,28) node [anchor=north west][inner sep=0.75pt]   [align=left] {3};
% Text Node
\draw (165,110) node [anchor=north west][inner sep=0.75pt]   [align=left] {4};
% Text Node
\draw (95,30) node [anchor=north west][inner sep=0.75pt]   [align=left] {i};
% Text Node
\draw (95,110) node [anchor=north west][inner sep=0.75pt]   [align=left] {j};
% Text Node
\draw (146,28) node [anchor=north west][inner sep=0.75pt]   [align=left] {i};
% Text Node
\draw (143,109) node [anchor=north west][inner sep=0.75pt]   [align=left] {j};
\end{tikzpicture}
}  \ \ \ \ \ \ \ \
\subfloat[]{
\tikzset{every picture/.style={line width=1pt}} %set default line width to 0.75pt       
\begin{tikzpicture}[x=0.75pt,y=0.75pt,yscale=-1,xscale=1]
 

%Straight Lines [id:da8952577887886422] 
\draw    (120.99,79.59) -- (99.61,58.21) ;
\draw [shift={(98.2,56.8)}, rotate = 45] [color={rgb, 255:red, 0; green, 0; blue, 0 }  ][line width=0.75]    (10.93,-4.9) .. controls (6.95,-2.3) and (3.31,-0.67) .. (0,0) .. controls (3.31,0.67) and (6.95,2.3) .. (10.93,4.9)   ;
%Straight Lines [id:da41451408396593414] 
\draw    (84.2,42.8) -- (98.2,56.8) ;

%Straight Lines [id:da06431575564936653] 
\draw    (120.65,79.13) -- (99.28,100.51) ;
\draw [shift={(97.87,101.92)}, rotate = 315] [color={rgb, 255:red, 0; green, 0; blue, 0 }  ][line width=0.75]    (10.93,-4.9) .. controls (6.95,-2.3) and (3.31,-0.67) .. (0,0) .. controls (3.31,0.67) and (6.95,2.3) .. (10.93,4.9)   ;
%Straight Lines [id:da16889697375559964] 
\draw    (83.87,115.92) -- (97.87,101.92) ;

%Straight Lines [id:da1644669094129687] 
\draw    (216.65,117.25) -- (195.28,95.88) ;
\draw [shift={(193.87,94.47)}, rotate = 45] [color={rgb, 255:red, 0; green, 0; blue, 0 }  ][line width=0.75]    (10.93,-4.9) .. controls (6.95,-2.3) and (3.31,-0.67) .. (0,0) .. controls (3.31,0.67) and (6.95,2.3) .. (10.93,4.9)   ;
%Straight Lines [id:da2984933625391206] 
\draw    (179.87,80.47) -- (193.87,94.47) ;

%Straight Lines [id:da7589691045098796] 
\draw    (216.99,43.47) -- (195.61,64.84) ;
\draw [shift={(194.2,66.25)}, rotate = 315] [color={rgb, 255:red, 0; green, 0; blue, 0 }  ][line width=0.75]    (10.93,-4.9) .. controls (6.95,-2.3) and (3.31,-0.67) .. (0,0) .. controls (3.31,0.67) and (6.95,2.3) .. (10.93,4.9)   ;
%Straight Lines [id:da3022498007196087] 
\draw    (180.2,80.25) -- (194.2,66.25) ;

%Shape: Ellipse [id:dp3579924082553517] 
\draw   (120.99,81.71) .. controls (120.99,70.69) and (134.24,61.75) .. (150.59,61.75) .. controls (166.94,61.75) and (180.2,70.69) .. (180.2,81.71) .. controls (180.2,92.73) and (166.94,101.67) .. (150.59,101.67) .. controls (134.24,101.67) and (120.99,92.73) .. (120.99,81.71) -- cycle ;
%Straight Lines [id:da24563613243517846] 
\draw    (150.59,61.75) -- (143.78,62.37) ;
\draw [shift={(141.79,62.55)}, rotate = 354.81] [color={rgb, 255:red, 0; green, 0; blue, 0 }  ][line width=0.75]    (10.93,-4.9) .. controls (6.95,-2.3) and (3.31,-0.67) .. (0,0) .. controls (3.31,0.67) and (6.95,2.3) .. (10.93,4.9)   ;
%Straight Lines [id:da4328176872479478] 
\draw    (150.59,101.67) -- (146.93,101.45) ;
\draw [shift={(144.94,101.33)}, rotate = 3.37] [color={rgb, 255:red, 0; green, 0; blue, 0 }  ][line width=0.75]    (10.93,-4.9) .. controls (6.95,-2.3) and (3.31,-0.67) .. (0,0) .. controls (3.31,0.67) and (6.95,2.3) .. (10.93,4.9)   ;

% Text Node
\draw (63,30) node [anchor=north west][inner sep=0.75pt]   [align=left] {1};
% Text Node
\draw (62,111) node [anchor=north west][inner sep=0.75pt]   [align=left] {2};
% Text Node
\draw (224,29) node [anchor=north west][inner sep=0.75pt]   [align=left] {3};
% Text Node
\draw (224,111) node [anchor=north west][inner sep=0.75pt]   [align=left] {4};
% Text Node
\draw (95,30) node [anchor=north west][inner sep=0.75pt]   [align=left] {i};
% Text Node
\draw (95,110) node [anchor=north west][inner sep=0.75pt]   [align=left] {j};
% Text Node
\draw (205,29) node [anchor=north west][inner sep=0.75pt]   [align=left] {i};
% Text Node
\draw (202,110) node [anchor=north west][inner sep=0.75pt]   [align=left] {j};
% Text Node
\draw (141,40) node [anchor=north west][inner sep=0.75pt]   [align=left] {i};
% Text Node
\draw (138.94,110.33) node [anchor=north west][inner sep=0.75pt]   [align=left] {j};
% Text Node
\draw (156,41) node [anchor=north west][inner sep=0.75pt]   [align=left] {5};
% Text Node
\draw (153,111) node [anchor=north west][inner sep=0.75pt]   [align=left] {6};

\end{tikzpicture}
} \ \ \ \ \ \ \ \
\subfloat[]{

\tikzset{every picture/.style={line width=1pt}} %set default line width to 0.75pt        

\begin{tikzpicture}[x=0.75pt,y=0.75pt,yscale=-1,xscale=1]

%Straight Lines [id:da6536284336590911] 
\draw    (120.99,79.59) -- (99.61,58.21) ;
\draw [shift={(98.2,56.8)}, rotate = 45] [color={rgb, 255:red, 0; green, 0; blue, 0 }  ][line width=0.75]    (10.93,-4.9) .. controls (6.95,-2.3) and (3.31,-0.67) .. (0,0) .. controls (3.31,0.67) and (6.95,2.3) .. (10.93,4.9)   ;
%Straight Lines [id:da9982534738877924] 
\draw    (84.2,42.8) -- (98.2,56.8) ;

%Straight Lines [id:da1798845392114402] 
\draw    (84.87,115.92) -- (106.24,94.55) ;
\draw [shift={(107.65,93.13)}, rotate = 135] [color={rgb, 255:red, 0; green, 0; blue, 0 }  ][line width=0.75]    (10.93,-4.9) .. controls (6.95,-2.3) and (3.31,-0.67) .. (0,0) .. controls (3.31,0.67) and (6.95,2.3) .. (10.93,4.9)   ;
%Straight Lines [id:da4073203382472357] 
\draw    (121.65,79.13) -- (107.65,93.13) ;

%Straight Lines [id:da3010890665331972] 
\draw    (216.65,117.25) -- (195.28,95.88) ;
\draw [shift={(193.87,94.47)}, rotate = 45] [color={rgb, 255:red, 0; green, 0; blue, 0 }  ][line width=0.75]    (10.93,-4.9) .. controls (6.95,-2.3) and (3.31,-0.67) .. (0,0) .. controls (3.31,0.67) and (6.95,2.3) .. (10.93,4.9)   ;
%Straight Lines [id:da07001653100141458] 
\draw    (179.87,80.47) -- (193.87,94.47) ;

%Shape: Ellipse [id:dp37207364622839356] 
\draw   (120.99,81.71) .. controls (120.99,70.69) and (134.24,61.75) .. (150.59,61.75) .. controls (166.94,61.75) and (180.2,70.69) .. (180.2,81.71) .. controls (180.2,92.73) and (166.94,101.67) .. (150.59,101.67) .. controls (134.24,101.67) and (120.99,92.73) .. (120.99,81.71) -- cycle ;
%Straight Lines [id:da7816491313522526] 
\draw    (156.11,62.11) -- (152.59,61.88) ;
\draw [shift={(152.59,61.88)}, rotate = 183.71] [color={rgb, 255:red, 0; green, 0; blue, 0 }  ][line width=0.75]    (10.93,-4.9) .. controls (6.95,-2.3) and (3.31,-0.67) .. (0,0) .. controls (3.31,0.67) and (6.95,2.3) .. (10.93,4.9)   ;
%Straight Lines [id:da6992984700470459] 
\draw    (150.59,101.67) -- (146.93,101.45) ;
\draw [shift={(144.94,101.33)}, rotate = 3.37] [color={rgb, 255:red, 0; green, 0; blue, 0 }  ][line width=0.75]    (10.93,-4.9) .. controls (6.95,-2.3) and (3.31,-0.67) .. (0,0) .. controls (3.31,0.67) and (6.95,2.3) .. (10.93,4.9)   ;
%Straight Lines [id:da8812714453372484] 
\draw    (179.87,80.92) -- (201.24,59.55) ;
\draw [shift={(202.65,58.13)}, rotate = 135] [color={rgb, 255:red, 0; green, 0; blue, 0 }  ][line width=0.75]    (10.93,-4.9) .. controls (6.95,-2.3) and (3.31,-0.67) .. (0,0) .. controls (3.31,0.67) and (6.95,2.3) .. (10.93,4.9)   ;
%Straight Lines [id:da5343707307385305] 
\draw    (216.65,44.13) -- (202.65,58.13) ;

% Text Node
\draw (63,30) node [anchor=north west][inner sep=0.75pt]   [align=left] {1};
% Text Node
\draw (62,111) node [anchor=north west][inner sep=0.75pt]   [align=left] {3};
% Text Node
\draw (224,29) node [anchor=north west][inner sep=0.75pt]   [align=left] {2};
% Text Node
\draw (224,111) node [anchor=north west][inner sep=0.75pt]   [align=left] {4};
% Text Node
\draw (95,30) node [anchor=north west][inner sep=0.75pt]   [align=left] {i};
% Text Node
\draw (95,110) node [anchor=north west][inner sep=0.75pt]   [align=left] {i};
% Text Node
\draw (205,29) node [anchor=north west][inner sep=0.75pt]   [align=left] {j};
% Text Node
\draw (202,110) node [anchor=north west][inner sep=0.75pt]   [align=left] {j};
% Text Node
\draw (141,40) node [anchor=north west][inner sep=0.75pt]   [align=left] {k};
% Text Node
\draw (138.94,110.33) node [anchor=north west][inner sep=0.75pt]   [align=left] {k};
% Text Node
\draw (156,41) node [anchor=north west][inner sep=0.75pt]   [align=left] {6};
% Text Node
\draw (153,111) node [anchor=north west][inner sep=0.75pt]   [align=left] {5};
\end{tikzpicture}
}
\caption{(a) Tree-level and (b) one-loop Feynman diagrams contributing to the kinetic equation.}\label{tree}
\end{figure}

The standard wave kinetic equation for the Hamiltonian (\ref{H1}) with one field was given in (\ref{WKE}). This is valid at leading order in the nonlinearity $\lam$. The kinetic equation at next-to-leading order in $\lam$ is:\cite{RS1, RS2, RSSS, Schubring} 
\be
\frac{\d n_1}{\d t} = 16\pi \text{Re} \sum_{p_2,p_3, p_4} \delta(\o_{p_1p_2; p_3 p_4}) \lam_{p_1p_2p_3p_4}^2  \prod_{i=1}^4 n_i\, \Big( \frac{1}{n_1} {+} \frac{1}{n_2}{-}\frac{1}{n_3} {-} \frac{1}{n_4} \Big)\\
\(1 + 2\mL_+ + 8\mL_-\)~,
\ee
where we have taken the coupling to be real, momentum conservation $p_1{+}p_2 = p_3{+}p_4$ is implied, and,
\be \label{A2}
\mL_+ =  2 \sum_{p_5}  \frac{\lam_{p_1p_2p_5p_6}\lam_{p_5p_6p_3p_4}}{\lam_{p_1p_2p_3p_4}}\frac{n_{p_5} {+} n_{p_6}}{\o_{p_1 p_2;p_5 p_6}{+}i\eps}~, \ \ \ \ \ \mL_- = 2\sum_{p_5}  \frac{\lam_{p_1p_6p_3p_5}\lam_{p_2p_5p_4p_6}}{\lam_{p_1p_2p_3p_4}} \frac{n_{p_6} {-} n_{p_5}}{\o_{p_1 p_6;p_3 p_5}{+}i\eps}~,
\ee
where, via momentum conservation, $p_6 = p_1{+}p_2{-}p_5$ (and the frequency is really the renormalized frequency). The three terms correspond to the  three Feynman diagrams shown in Fig.~\ref{tree}, where for the moment one should ignore the $i,j,k$ indices. In principle, one can compute the kinetic equation to arbitrary order in $\lam$ \cite{RSSS}, but the higher the order, the more terms there are and the more unwieldy the expression.

%%%%%%%%
%%%%% START FIGURE
%%%%%%%%
%%%%% End FIGURE

Introducing $N \gg 1$ fields, so that the Hamiltonian is given by (\ref{HN}), gives a drastic simplification. The diagrams now have indices which label the fields. Since each interaction vertex introduces a factor of $1/N$, in order for a higher order diagram involving loops to not  be suppressed at large $N$, it must have an index sum over all fields in the loop to compensate for this factor. For instance, Fig.~\ref{tree}(c) survives the large $N$ limit (due to the sum over $k$) but Fig.~\ref{tree}(b) does not. Likewise, at higher loop order the only diagrams are the bubble diagrams shown earlier in Fig.~\ref{bubblesum}. 
The sum of all bubble diagrams gives the large $N$ kinetic equation \cite{RS}, 
\be \label{WKE2}
\frac{\d n_1}{\d t} = \frac{8\pi}{N}\!\!\! \sum_{p_2,p_3, p_4}  |\Lambda_{p_1 p_2p_3p_4}|^2  \prod_{i=1}^4 n_i\, \Big( \frac{1}{n_1} {+} \frac{1}{n_2}{-}\frac{1}{n_3} {-} \frac{1}{n_4} \Big)\delta(\o_{p_1}{+}\o_{p_2}{-} \o_{p_3}{-}\o_{ p_4})~,
\ee
where the ``effective coupling'' $\Lambda_{p_1 p_2p_3p_4}$ is given by an intuitively clear expression, representing a geometric sum where each loop adds a factor of $\mL$,
\be \label{Lam}
\Lambda = \lam(1 - \mL \lam)^{-1}
\ee
where $(1 - \mL \lam)^{-1} = 1 + \mL\lam + \mL\lam \mL\lam + \ldots$, and we should think of this as a matrix product with indices that live in momentum space. Namely, defining $q = p_1{-}p_3$, the coupling $\lam_{p_1p_2p_3 p_4}$ is represented by $\lam_{14} \equiv \lam_{p_1, p_4{-}q,  p_1{-}q, p_4}$, where we made use of momentum conservation, and $\mL$ is a diagonal matrix, 
\be
\mL_{ab} = \delta_{a b} \frac{2(n_{a-q} {-}n_a)}{\o_{p_1}{-}\o_{p_3}{+}\o_{a-q}{-}\o_a {+}i\eps}~.
\ee 
Thus, explicitly, $(\lam \mL \lam)_{14} = \sum_{a,b} \lam_{1 a} \mL_{a b} \lam_{b4} = \lam_{p_1 p_2 p_3 p_4} \mL_-$, where $\mL_-$ was given in (\ref{A2}). Note that $\mL$ depends on $\o_{p_1}{-}\o_{p_3}$ and $q=p_1{-} p_3$, which are not part of the matrix indices. For a general $\lam$, one can not write $\Lambda$ any more explicitly --- in effect, it is given by the solution of (\ref{Lam}), which is an integral equation. In the special case that the interaction has product factorization: $\lam_{p_1 p_2 p_3 p_4}= \sqrt{\lam_{p_1 p_3} \lam_{p_2 p_4}}$ \cite{FR}, the expression is simple, 
\be
\Lambda_{p_1 p_2p_3p_4} = \frac{\lam_{p_1 p_2 p_3 p_4}}{1- \mL_-}~.
\ee
For a constant interaction, $\lam_{p_1 p_2 p_3 p_4} = \lam$, this reduces to the large $N$ kinetic equation found in \cite{Walz:2017ffj,bergesGasenzerScheppach2010, berges2002,bergesRothkopfSchmidt2008}. 

For the strongly local interactions discussed in the main body,  in the relevant one loop term $\lam_{p_1 p_2 p_3 p_4} \mL_-$ (\ref{A2}) the integral over $p_5$ will be localized in the vicinity of $p_1$. Taylor  expanding  $n_{p_6}- n_{p_5}$ gives $\o \frac{\d n}{\d \o}$. The remaining integral in the vicinity of $p_5$ is some constant $c$. We assume each loop gives the same factor, so we have a geometric sum. The result is the kinetic equation (\ref{KE}) given in the main body. Note that $\mL_-$ has both a real and imaginary part, so $c$ is in general complex.

\section{Strong wave turbulence from dimensional analysis} \label{apC}

In this appendix we reproduce, on the basis of dimensional analysis,   the strong turbulence scaling found in Sec.~\ref{sec3}:  $n_k \sim k^{- d- 2\al}$ for an energy cascade and $n_k \sim k^{- d- \al}$ for a particle number cascade, where  $\al$ is the scaling exponent of the frequency, $\o_k \sim k^{\al}$.

First, recall the famous Kolmogorov $5/3$ scaling law: assuming the only dimensionful parameter is the density $\rho$, having dimensions $M/L^d$, the energy flux $P$ has dimensions of energy per unit volume per unit time: $P\sim \rho L^2/T^3$, while the energy density $\varepsilon_k$ in $k$ space ($k = 1/L$) is energy per unit $k$ per unit volume, $\varepsilon_k \sim  \rho L^3/T^2$. Matching dimensions requires $\varepsilon_k \sim k^{-5/3} P^{2/3} \rho^{1/3}$. 

In weak wave turbulence there is a second dimensionful parameter, which comes from the dispersion relation $\o_k$, having dimensions $1/T$. For instance, for gravity waves $\o_k \sim \sqrt{g k}$, where $g$ is the gravitational constant. To obtain the Kolmogorov-Zakharov spectrum one makes use of dynamics: the weak wave kinetic equation, which gives the time scale $t_k$: $n_k/t_k \sim n_k^3 k^{2d} \lam_k^2/\o_k$.  Upon inserting $t_k$ into the flux $P\sim \o_k n_k k^d/t_k\sim k^{3(d-\g) + 2\beta}$, and taking $P$ to be a constant, one recovers the KZ solution, $n_k\sim k^{- (d+\frac{2}{3}\beta)}$. In the case of large $N$ strong wave turbulence, one can use the large $N$ kinetic equation (\ref{WKE2}, \ref{Lam}) to establish $t_k$. Namely, for strong nonlinearity one drops the $1$ relative to $\mL \lam$ in (\ref{Lam}) and thus finds $1/t_k \sim \o_k$. Inserting $t_k$ into the flux $P\sim \o_k n_k k^d/t_k$ gives $n_k \sim k^{- (d+2\al)}$, the strong turbulence solution found in (\ref{S1}). 

For strong wave turbulence more generally, we do not know the kinetic equation. However, if the strength of the nonlinear interaction $\lam_{p_1 p_2 p_3 p_4}$ is taken to be arbitrarily large, then $\lam_{p_1 p_2 p_3 p_4}$ can not enter into determining the spectrum. Therefore, like in Kolmogorov scaling, we are back to one dimensionful parameter. However, now it is not density but rather  $g$ (for gravity waves) or, more generally, a dimensionful constant $\mathrm{g}$ with dimension $L/T^{1/\al}$ for the dispersion relation $\o_k \sim (k \mathrm{g})^{\al}$. Relating energy density $\varepsilon_k \sim E/L^{d-1}$ and flux $P\sim E/(L^d T)$ gives $\varepsilon_k/P\sim L T \sim \mathrm{g}^{-\al} k^{-1-\al}$ and hence $n_k \sim \varepsilon_k/(k^{d-1} \o_k) \sim P \mathrm{g}^{-\al} k^{-d-2\al}$.  A similar argument gives the scaling for a particle flux cascade, with nonzero $Q$ rather than $P$. This dimensional analysis argument may explain why numerical results in \cite{Nowak:2011sk} saw this scaling for $\al =2 , \beta =0 $ for $N=1$ and in \cite{Berges:2010ez} for $\al =1 , \beta = -2$ for $N=4$. 

It will be interesting to study to what extent this scaling is present in physical examples of strong wave turbulence. If this scaling is actually realized -- in the sense that the assumption that the pumping and dissipation scales don't matter  is valid -- is potentially system specific.

\section{ Numerical method} \label{appendixB}
\begin{figure}[]
	\includegraphics[width=0.5\textwidth]{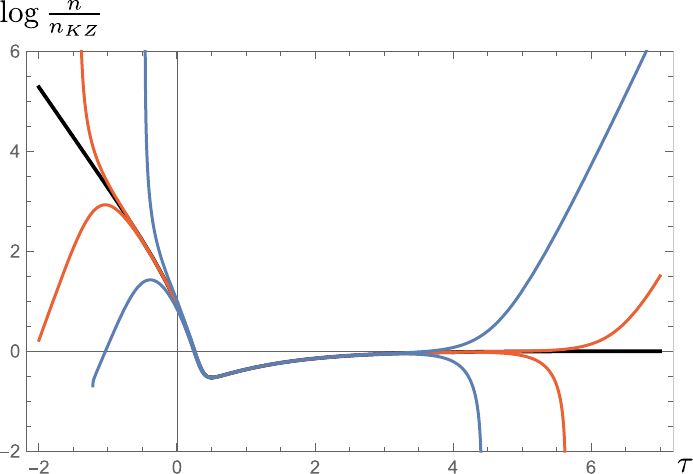}	\centering
	\caption{Illustration of the shooting method at $\tau_0=1$. The parameters are as in Fig~\ref{fig1}(b). The blue curves involve initial conditions $\dot{v}(1)=5.3432$ and $v(1)=7.8374$ and $7.8375$ for top and bottom, respectively. The red curves involve initial conditions $\dot{v}(1)=5.343167$ and $v(1)=7.837440, 7.837441$, respectively. In black, a piecewise solution also making use of the initial values $v(-1)=0.02809349$, $\dot{v}(-1)=0.08560496,$ and $v(4)=111.48895$, $\dot{v}(4)=109.18942$ is shown.}\label{figNumMethod}
\end{figure}

The main numerical problem involves finding solutions of equation \eqref{eqMain} that interpolate between various scaling solutions at $\tau=\pm \infty$. This was done using a variant of the shooting method for boundary value problems. The initial values $v_0=v(\tau_0), \dot{v}_0=\dot{v}(\tau_0)$ for some $\tau_0$ are chosen, and the differential equation is solved using standard methods for initial value problems. Then the initial values $v_0, \dot{v}_0$ are adjusted to increasingly greater precision in order to balance between different classes of asymptotic behavior.

For example, consider the strong to KZ curve in Fig~\ref{fig1}(b). To five digits, the initial value of the curve at $\tau_0=1$ is $v_0= 7.8374, \dot{v}_0= 5.3432$. The solution to this initial value problem displays thermal behavior in the UV and $\log n$ going to positive infinity at a finite value of $\tau$ in the IR. If the last digit of $v_0$ is increased to 7.8375, then the asymptotic behavior changes on both sides, with $\log n$ going to negative infinity in both the UV and IR (the blue curves in Fig.~\ref{figNumMethod}). At the next step of the algorithm, an additional digit is introduced, and the solution is increased to a new threshold at $v_0=7.83744$. If upon adjusting $v_0$ we ever reach the situation where $\log n$ goes to negative infinity on only one side, then the initial slope $\dot{v}_0$ is adjusted until it once again goes to negative infinity on both sides or on neither side. After adjusting the slope, the corresponding red curves in Fig.~\ref{figNumMethod} display strong and KZ scaling over an increased range.

This procedure allows us to specify the initial conditions with increasing precision, but eventually the solution at points distant from $\tau_0$ may be affected by numerical error in the method for solving the initial value problem. Instead, we may find solutions over a wider range by constructing piecewise solutions involving multiple initial conditions, as in the black curve of Fig \ref{figNumMethod}. An important consistency check on the piecewise solutions is that the initial conditions for each interval fall within the error range predicted by the adjacent intervals.

\section{ Additional cases} \label{apD}
\subsection*{Complex values of $c\, \lambda$}
\begin{figure}[]
	\includegraphics[width=0.5\textwidth]{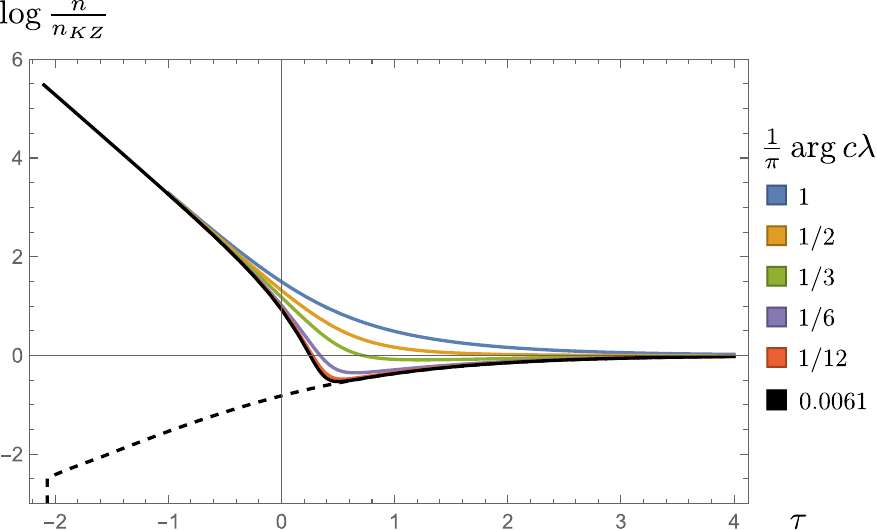}	\centering
	\caption{Stationary solutions of the strongly local large $N$ kinetic equation for complex values of $c \lam$, see \eqref{26} or (\ref{eqMainComplex}). The parameters are as in Fig.~\ref{fig1}(b). The strong to KZ scaling curves for $\arg c\lambda=0$ and $0.0061\pi$ (black) are indistinguishable at this resolution. The long-lived Phillips solution for $0.0061\pi$ is shown as a dashed line.}\label{figComplexC}
\end{figure}
We have mostly been considering the case in which the  parameter $c\, \lambda$ introduced in the strongly local approximation is real and positive. The situation in which $c\, \lambda$ is real and negative was  shown in the dotted line solutions of Fig.~\ref{fig1}. As mentioned in Appendix~\ref{apA}, $c$ is naturally complex ($\lam$ can be made complex too if one wishes).  In general, we may write $c\, \lambda = |c\, \lambda|e^{i\phi}$, and after defining $v$ and $\epsilon$ in terms of $|c\, \lambda|$, Eq.~\ref{eqMain} is generalized to,
\begin{align}\label{eqMainComplex}
	\ddot{v} + (2\xi{-}1) \dot{v} +\xi(\xi{-}1) v = \epsilon^3\(v^4 +(\xi v + \dot v)^2-2\cos\phi\,v^2(\xi v + \dot v)\).
\end{align}

The solutions interpolating between strong turbulence and KZ scaling for various values of $\phi= \arg c\lambda$ are shown in Fig.~\ref{figComplexC}.
Unless $\cos\phi=1$, it is clear that there is no true Phillips solution at large enough $\epsilon$, but for intermediate $\epsilon$ the situation is more subtle. For $1>\cos\phi>\sqrt{3}/2$, there is a finite range of $\epsilon$ such that there are three fixed points -- like  in the phase portrait of Fig.~\ref{figPhasePlot} -- and once again points $a$ and $b$ are associated with Phillips scaling. However, now there is some maximum $\epsilon$ where points $a$ and $b$ merge and disappear, and the Phillips solution does not extend arbitrarily far into the strong coupling regime.

In Fig.~\ref{figComplexC} the value $\phi=0.0061\pi$ depicted in black is chosen so that the points $a$ and $b$ disappear at $\tau={-}2$. A solution interpolating between Phillips and KZ scaling for $\tau>-2$ is shown as a dashed line. For $\tau<-2$, the curve abruptly decays to negative infinity, corresponding to $n$ vanishing just below $\tau=-2.07$.

\subsection*{Phillips exponent less than thermal}
\begin{figure}[]
\subfloat[]{	\includegraphics[width=0.4\textwidth]{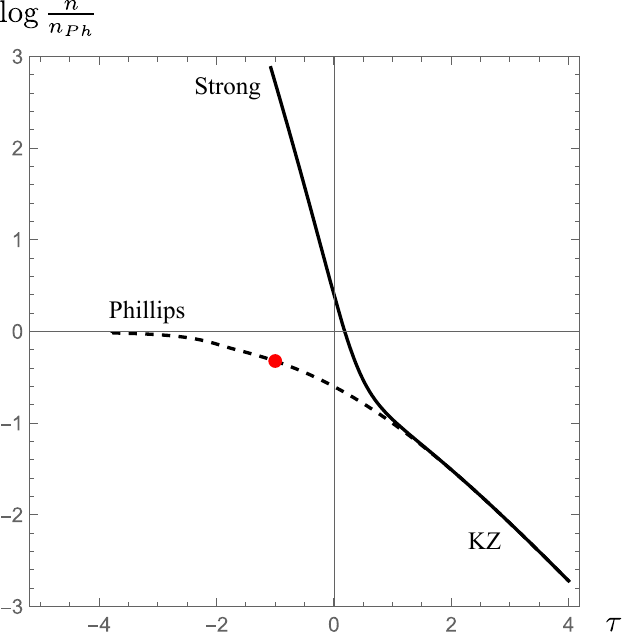}} \ \ \ \ \ \ \ \ \ \ \  
\subfloat[]{
	\includegraphics[width=0.41\textwidth]{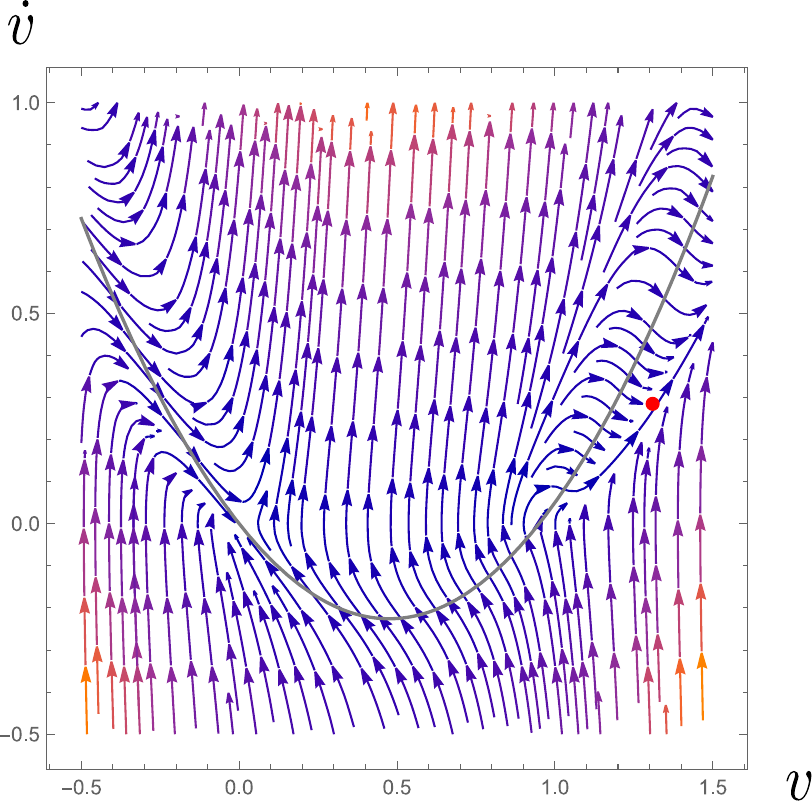}}		\centering
	\caption{ A solution displaying long-lived Phillips behavior in the IR for $\xi{=}0.95$ ($d{=}2, \alpha{=}2, \beta{=}1.9, \tilde{P}{\equiv} c^3\omega_0^{3\kappa}\lambda P{=}3/2$). (a) is a plot of $\log (n/n_{Ph})$ vs $\tau$ where we normalized by the pure Phillips scaling $n_{Ph}$ (\ref{38}) , and (b) is a phase portrait at $\tau=-1$. The point on the solution $v(-1)=1.30963$, $\dot{v}(-1)=0.284960$ is plotted in red. As $\tau$ decreases, the solution survives as long as it stays in the region near the parabola $\dot{v}=v(v{-}\xi)$ plotted in gray in (b), but at $\tau\approx -3.78$ it falls off and diverges.}\label{figSmallNu}
\end{figure}

So far in discussing the Phillips solution we have implicitly assumed that the exponent $\xi$ (defined in (\ref{44})) is greater than the thermal exponent of $1$. For $\xi<1$, there are no longer Phillips solutions that extend arbitrarily far into the IR.\footnote{In this section we focus on the case where $\epsilon$ is large in the IR. If it is instead large in the UV that implies that the strong scaling exponent $d/\alpha + 2$ is less than the Phillips exponent $\xi\leq 1$, which is impossible if we take the physical conditions $d,\alpha>0$. } This may be seen from the fixed point equation arising from \eqref{eqMain},
\begin{align}
\xi(\xi-1)=\epsilon^3(v-\xi )^2v, \label{eqMainFixedPoint}
\end{align}
which has no physical solution with $\epsilon, v>0$ if the left hand side is negative. It may also be seen from the asymptotic Phillips solution in \eqref{47}. When $\xi=\zeta+\kappa<1$ the correction term becomes imaginary.

However, if $\xi$ is slightly less than $1$, there may still be long-lived solutions that display Phillips scaling for some finite range, see Fig.~\ref{figSmallNu}. This is possible if the solutions stay in the vicinity of the parabola $v^2-(\xi v+\dot{v})=0$ in phase space. This parabola is equivalent to the first-order differential equation associated to the Phillips behavior \eqref{eqAsympPhillips}. There is a relatively slow moving region in phase space around this parabola, which becomes thinner as $\epsilon$ increases. As long as the solution remains in the slow moving region near $v\approx \xi$, it displays Phillips scaling. However, since for $\xi<1$ there is no fixed point to stop it, eventually in the IR the solution falls to the region of large negative $\dot{v}$ and diverges.

In the special case of $\xi = 1$, the exponents for Phillips and thermal scaling are identical. The fixed point equation \eqref{eqMainFixedPoint} has an exact fixed point at $v=\xi, \dot{v}=0$ for all $\epsilon$, and it is possible to get pure Phillips behavior for all $\tau$, even in the weak regime. There is also a solution that interpolates between Phillips and KZ scaling, much as in the  $\xi>1$ case. A qualitative difference from the $\xi>1$ case shown in Fig.~\ref{fig1}(b) is that the family of additional Phillips to KZ solutions existing for all $\tau$ shown in purple no longer exists. This may perhaps be understood from the perspective of the fixed point equation \eqref{eqMainFixedPoint}, where fixed points $a$ and $b$ in Fig.~\ref{figPhasePlot} collapse to a single double root at $\xi=1$.

\subsection*{Kolmogorov-Zakharov exponent less than thermal}
\begin{figure}[]
	\includegraphics[width=0.45\textwidth]{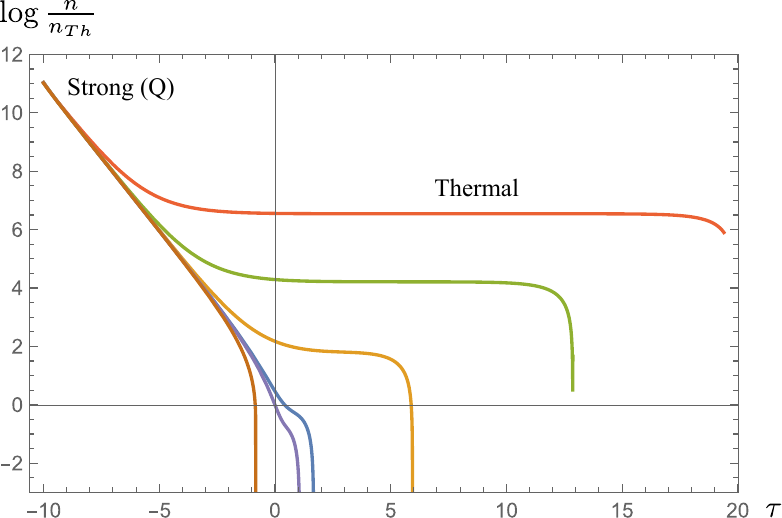}	\centering
	\caption{Solutions displaying strong scaling in the IR for $\zeta{=}2/3$ ($d{=}2, \alpha{=}2, \beta{=}0, P{=}0, \epsilon^3(0){=}3/2$). The vertical axis is scaled by $n_{Th}\equiv (c\lambda)^{-1}\omega_0^{1-\xi}\omega^{-1}$ such that solutions with thermal scaling have zero slope. All strong scaling solutions decay in the UV, but there can be long-lived thermal behavior.}\label{figSmallK}
\end{figure}

If, in addition, the  exponent $\zeta$ (defined in (\ref{41})) is less than the thermal exponent of $1$, there is neither Phillips nor KZ scaling behavior. The disappearance of the KZ scaling behavior for $\zeta\leq1$ is clear from the vanishing of $u$ at $\zeta=1$ in the asymptotic expression \eqref{47}, and it has been discussed for the weak wave turbulence equation \eqref{eqAsympWeak} in \cite{Naz}. However, there may still be strong turbulence scaling in the IR, and several such solutions are plotted in Fig.~\ref{figSmallK}. The strong turbulence solutions may interpolate to long-lived thermal behavior, but all solutions eventually decay. Precisely at $\zeta=1$ the thermal exponent and KZ exponent are identical, so the long-lived solutions may be alternatively understood as KZ solutions with logarithmic modifications \cite{DYACHENKO}.

We note in passing that although there are no thermal solutions extending arbitrarily far into the UV for $\zeta\leq1$, there are solutions such that $n$ is asymptotically constant in the UV. These asymptotically constant solutions decay in the IR, and can not be reached from the strong scaling solution. These solutions appear already in the weak wave turbulence equation \eqref{eqAsympWeak}, and are equivalent to motion along the separatrix labeled $S$ in Fig.~6 and 9 in \cite{Naz}. Although they are more obvious in the case $\zeta \leq 1$, which has a dearth of solutions surviving in UV, these asymptotically constant solutions (and the associated separatrix in a phase portrait of \eqref{eqAsympWeak}) have nothing to do with the condition $\zeta \leq 1$ per se, and they exist also for $\zeta{>}1$.

\bibliographystyle{utphys}
%\bibliography{LargeNTurbulenceBib}

\end{document}